\documentclass[letterpaper]{article} 
\usepackage{aaai24}  
\usepackage{times}  
\usepackage{helvet}  
\usepackage{courier}  
\usepackage[hyphens]{url}  
\usepackage{graphicx} 
\usepackage{natbib}  
\usepackage{caption} 
\usepackage{algorithm}
\usepackage{algorithmicx}
\usepackage{amsmath,amssymb,amsfonts}
\usepackage{subfig}
\usepackage{algpseudocode}
\usepackage{upgreek}
\usepackage{enumerate}
\usepackage{color}
\usepackage{newfloat}
\usepackage{listings}
\DeclareCaptionStyle{ruled}{labelfont=normalfont,labelsep=colon,strut=off} 
\floatstyle{ruled}
\newfloat{listing}{tb}{lst}{}
\floatname{listing}{Listing}
\graphicspath{{figure/}}

\title{Model-Driven Deep Neural Network for Enhanced AoA Estimation Using 5G gNB}
\author{
Shengheng~Liu\textsuperscript{\rm 1, 2 *},
Xingkang~Li\textsuperscript{\rm 1},
Zihuan~Mao\textsuperscript{\rm 1, 2},
Peng~Liu\textsuperscript{\rm 2},
and Yongming~Huang\textsuperscript{\rm 1, 2 *}
}
\affiliations{
    \textsuperscript{\rm 1}National Mobile Communications Research Laboratory, Southeast University, Nanjing, China\\
    \textsuperscript{\rm 2}Purple Mountain Laboratories, Nanjing, China\\

  \{s.liu; huangym\}@seu.edu.cn
}

\usepackage{bibentry}

\begin{document}

\maketitle

\begin{abstract}
High-accuracy positioning has become a fundamental enabler for intelligent connected devices. Nevertheless, the present wireless networks still rely on model-driven approaches to achieve positioning functionality, which are susceptible to performance degradation in practical scenarios, primarily due to hardware impairments. Integrating artificial intelligence into the positioning framework presents a promising solution to revolutionize the accuracy and robustness of location-based services. In this study, we address this challenge by reformulating the problem of angle-of-arrival (AoA) estimation into image reconstruction of spatial spectrum. To this end, we design a model-driven deep neural network (MoD-DNN), which can automatically calibrate the angular-dependent phase error. The proposed MoD-DNN approach employs an iterative optimization scheme between a convolutional neural network and a sparse conjugate gradient algorithm. Simulation and experimental results are presented to demonstrate the effectiveness of the proposed method in enhancing spectrum calibration and AoA estimation.
\end{abstract}

\vspace{-0.8em}
\section{Introduction}

Development of the forthcoming wireless communications entails the exploration and integration of cutting-edge technologies and paradigms for diverse and emerging applications \cite{YOU2023Toward}. Among the essential components, high-accuracy positioning functionality assumes a critical role in guaranteeing a seamless user experience and accommodating various use cases such as emergency response, asset tracking, and location-based services \cite{Sun2021IDOL}. Despite being in its nascent stages, an escalating number of 5G mobile base stations (a.k.a. gNBs) have been deployed on a massive scale in both indoor and outdoor areas, offering an extraordinary platform for supporting positioning services \cite{Pai2019Advances}. As a result,  there is a renewed drive for the perception of the surrounding environment and the provision of location-aware services from the physical layer to the application layer.

In Release 16 of the 3GPP, which was finalized in 2021, several positioning schemes for 5G new radio (NR) have been introduced. These include Downlink-time-difference-of-arrival (DL-TDoA), uplink-time-difference-of-arrival (UL-TDoA), downlink-angle-of-departure (DL-AoD), uplink-angle-of-arrival (UL-AoA), and multi-round trip time (Multi-RTT) positioning. Among these schemes, UL-AoA has been
identified as more feasible for mitigating the multipath effect of the receive signal and enhancing the positioning performance, as it aligns with the trend towards increasing array size and bandwidth. However, purely model-driven AoA estimation algorithms usually suffer from hardware impairments that can affect the array response used in the estimation process \cite{pan2022efficient}.

By leveraging artificial intelligence (AI) algorithms, such as machine learning and deep neural networks (DNN), wireless networks can now dynamically learn and adapt to real-world conditions, compensating for hardware impairments and environmental variations \cite{Dai2022DeepAoANet,Lee2023Deep}. To address this issue, data-driven methods, such as DNN-based direction finding techniques, have surged exponential popularity \cite{Ghourchian2017Real,Mo2023Unified}. Nonetheless, the accuracy of such methods is heavily reliant on the availability of massive data sets, which hinders their translation to practical uses. The integration of neural network (NN) and model-driven methods \cite{Zhu2022EfficientMN, huang2023tdoa} provides inspiration for developing a model-driven deep learning method to mitigate the impact of hardware impairments and improve the AoA estimation performance.

Building on these recent advancements, this paper presents an innovative model-driven deep neural network (MoD-DNN) framework, featuring spectrum calibration capabilities to enhance AoA estimation. Simulation and experimental results demonstrate that the proposed framework yields superior AoA estimation performance compared to both purely model-driven and data-driven methods. To the best of our knowledge, this study marks the first AI-empowered positioning endeavor utilizing commodity 5G NR gNB. The key contributions of this paper include: (1) The problem formulation of an inverse problem from the coarray spatial spectrum (CSS) to a sparse solution of the over-completed AoA set; (2) Employing a one-dimensional convolutional neural network (1D-CNN) for CSS calibration, followed by a sparse conjugate gradient (SCG) algorithm for sparse AoA spectrum recovery; and (3) Proposing an iterative optimization strategy utilizing shared CNN coefficients between the CNN and SCG modules to establish the comprehensive MoD-DNN framework.

Notations: Lower (upper)-case bold characters are used to denote vectors (matrices), and the vectors are by default in column orientation.  ${\rm {E}}[\cdot]$ denotes the expected value of a discrete random variable. The superscripts $(\cdot)^{\mathsf T}$, and $(\cdot)^{\mathsf H}$ represent the transpose and conjugate transpose operators, respectively. Symbols $\odot$ stands for Hadamard product. $\jmath = \sqrt{-1}$ is the imaginary unit.

\vspace{-0.6em}
\section{Related Work}

Recently, the success of machine learning across a wide range of disciplines gave rise to NN-aided AoA estimation. With the continuous enhancement of computational power, NN have  widespread application in localization systems.

\subsubsection{End-to-End Learning}

Numerous studies utilize deep CNNs to directly map input signals to estimated AoA parameters in an end-to-end fashion \cite{Xuan2023End}. Augmenting the DNN architecture enables both offline and online learning for AoA estimation \cite{Huang2018Deep}, with multiple hidden layers enhancing recognition potential and sparse feature extraction within the angular domain \cite{Gehring2021Understanding}. However, their efficacy hinges on NN's generalization capacity \cite{Kotary2021End}, necessitating substantial training data and associated labels.

\subsubsection{Feature Learning}

Alternative methodologies incorporate a feature learning strategy \cite{Hou2019Learning}, yielding a subtle improvement in physical interpretability. By leveraging neural networks, specific input parameter features \cite{Naseri2022Machine} are acquired and refined, subsequently applied for AoA estimation, thereby reinforcing the linkage between input and estimated parameters with discernible physical significance. Nonetheless, these approaches, centered solely on training a single parameter \cite{Jiang2019Joint}, run the risk of omitting valuable information pertinent to other critical feature parameters.

\subsubsection{Model-Driven Learning}

Diverging from pure data-driven neural networks, certain methodologies integrate principled mathematical models with data-driven systems, reaping the strengths of both paradigms \cite{Zhou2020Deep, Zhu2020Bridging, fan2023fast}. Hybrid model-driven deep learning schemes have been advanced to synergize prior statistical models \cite{Merkofer2022Deep, HasanzadeZonuzy2021Model}. These hybrid methods strategically harness partial domain knowledge provided by mathematical structures tailored to specific problems, while also leveraging learning from limited data.

\begin{figure*}[t]
	\centering
	\includegraphics[width=0.99\linewidth,trim=40 20 30 30,clip]{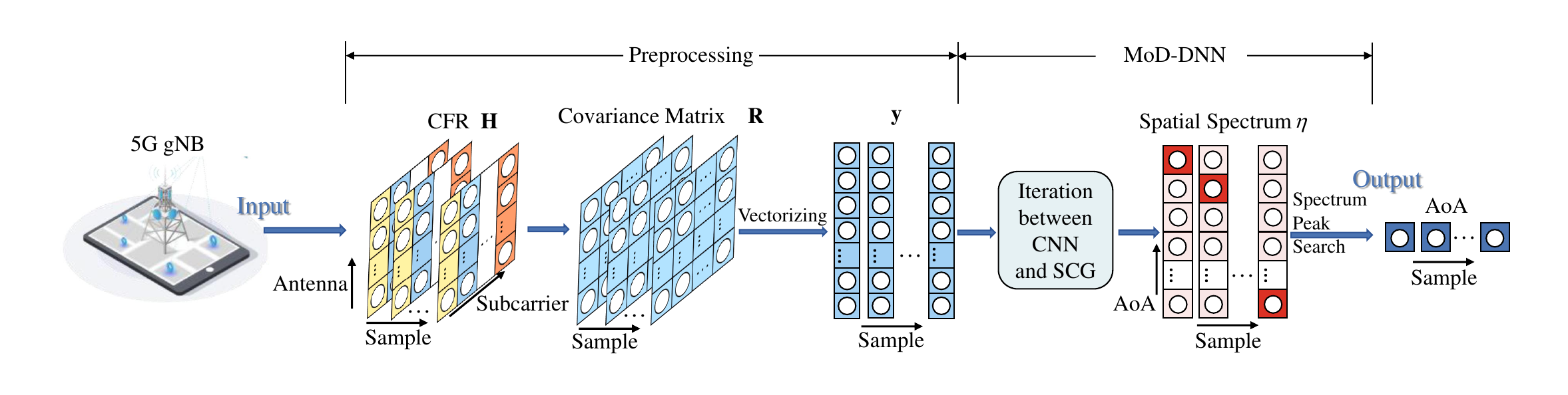}
\vspace{-1.2em}
	\caption{Framework of MoD-DNN for AoA estimation.}
\vspace{-0.8em}
	\label{framework}
\end{figure*}

\section{Problem Formulation}\label{sec:IdealSignal}

\subsubsection{Ideal Signal Model}

To begin with, we consider the ideal signal output in an uplink positioning scenario. Without loss of generality, the field of view is quantified using uniform discrete AoA set $\boldsymbol{\theta}=[\theta_1,\theta_2,\cdots,\theta_{L}]^{\mathsf{T}}$ with a size of $L$, the antenna array of 5G gNB is an $M$-element uniform linear array (ULA) with unit inter-element spacing equivalent to half wavelength. Assuming that the uniform frequency increment of the orthogonal frequency division multiplexing (OFDM) signal with $K$-subcarrier is much smaller than the central carrier frequency, the signal of the $k$-th subcarrier is expressed as
\vspace{-0.4em}
\begin{align}\label{signalmodel}
	\mathbf{x}_{k}(t) &= \sum\nolimits_{l=1}^{L}\mathbf{a}(\theta_l)s_{l,k}(t)+\mathbf{n}_{k}(t)\nonumber \\
	&=\mathbf{A}(\boldsymbol{\theta})\mathbf{s}_{k}(t)+\mathbf{n}_{k}(t), \quad t=1,\cdots,T,
\end{align}
where $s_{l,k}(t)$ represents the component on the $k$-th subcarrier of the sounding reference signal (SRS) transmitted by the UE locating at azimuth $\theta_l$. $\mathbf{n}_{k}(t)$ denotes the additive white Gaussian noise (AWGN). $\mathbf{a}(\theta_l)=[1,\cdots,{\rm{e}}^{{\jmath \pi  (M-1) \sin\theta_l}}]^{\rm{T}}$ and $\mathbf{A}(\boldsymbol{\theta})=[\mathbf{a}(\theta_1),\cdots,\mathbf{a}(\theta_L)]$ are the steering vector and  array manifold, respectively. Note that, the above ideal array manifold $\mathbf{A}$ is exactly determined by the azimuth of the UE and the ideal array configuration. By implementing fast Fourier transform (FFT), the channel state information (CSI) of the $k$-th subcarrier is derived as
\begin{align}\label{CSI}
	\mathbf{h}(k) = \mathbf{A}(\boldsymbol{\theta})\breve{\mathbf{s}}(k)+\breve{\mathbf{n}}(k),\quad k=1,\cdots,K,
\end{align}
where $\breve{\mathbf{s}}(k)$ and $\breve{\mathbf{n}}(k)$ are signal and noise in frequency domain. In this case, $\mathbf{h}(k)$ is equivalent to a single snapshot signal. As hardware impairments generally exist in real-world systems, leading to the mismatch of  the steering vectors $\tilde{\mathbf{a}}(\theta)$ in the real system and the ideal $\mathbf{a}(\theta)$, the performance of purely model-driven AoA estimators based on the ideal signal model usually degrade in practice.

\subsubsection{Effect of Hardware Impairments}

Several types of error are proposed to model the effect of hardware impairments, which are the element position error, gain and phase inconsistencies, and mutual coupling \cite{Pan2022In}. In most cases, the hardware impairments lead to phase error versus different AoAs. To precisely model the phase error into the signal model, the CSI-form receive signal of a practical system can be expressed as
\vspace{-0.4em}
\begin{align}\label{angular_dependent_error}
	\mathbf{h}(k) &= \sum\nolimits_{l=1}^{L}\boldsymbol{\gamma}(\theta_l)\odot\mathbf{a}(\theta_l)\breve{s}_l(k)+\breve{\mathbf{n}}(k)\nonumber \\ &=\mathbf{\Gamma}\odot\mathbf{A}(\boldsymbol{\theta})\breve{\mathbf{s}}(k)+\breve{\mathbf{n}}(k),
\end{align}
where $\boldsymbol{\gamma}(\theta_l)=[\gamma_{1}(\theta_l),\cdots,\gamma_{M}(\theta_l)]^{\mathsf{T}}$ and error matrix $\mathbf{\Gamma}=[\boldsymbol{\gamma}(\theta_1);\cdots;\boldsymbol{\gamma}(\theta_L)]$. The coefficient $\gamma_{m}(\theta_l)$ denotes the phase error of the $m$-th antenna for azimuth $\theta_l$. Thus, the practical steering vector $\tilde{\mathbf{a}}(\theta_l) = \boldsymbol{\gamma}(\theta_l)\odot\mathbf{a}(\theta_l)$, and practical array manifold $\tilde{\mathbf{A}}(\boldsymbol{\theta})=[\tilde{\mathbf{a}}(\theta_1),\cdots,\tilde{\mathbf{a}}(\theta_L)]$. Formula \eqref{angular_dependent_error} reveals that the hardware impairment results in angular-dependent and nonlinear phase error in practical system.

\section{Network Framework}

To overcome the aforementioned influence, we propose a MoD-DNN framework with the ability of spectrum calibration to enhance AoA estimation performance. A sketch of the proposed framework is illustrated in Figure~\ref{framework}. In the preprocessing module, we first reformulate the input signal to a co-array signal form. Then, the MoD-DNN reconstruct the sparse spatial spectrum via iterative optimization between 1D-CNN and SCG. In the following, we respectively introduce those modules of the proposed framework in detail.
\begin{figure*}[t]
	\centering
	\subfloat[]
	{\includegraphics[width=0.56\linewidth]{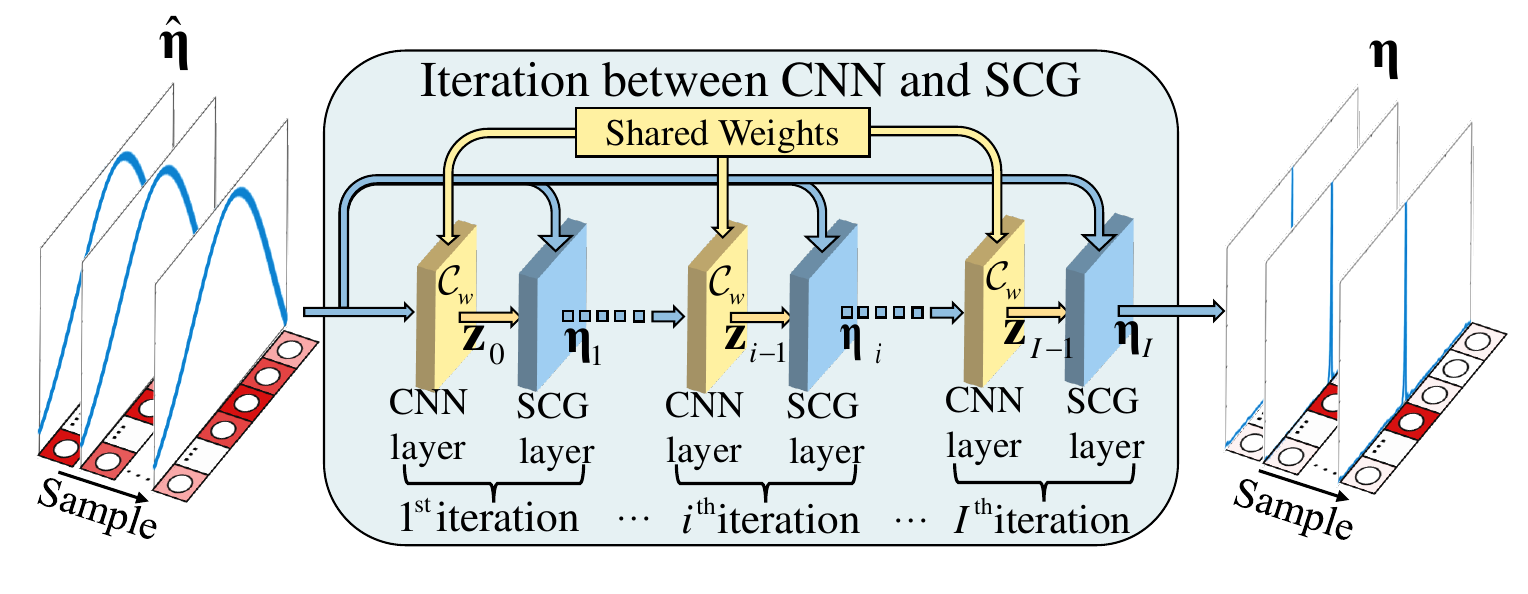}}
	\hfil
	\centering
	\subfloat[]
	{\includegraphics[width=0.42\linewidth]{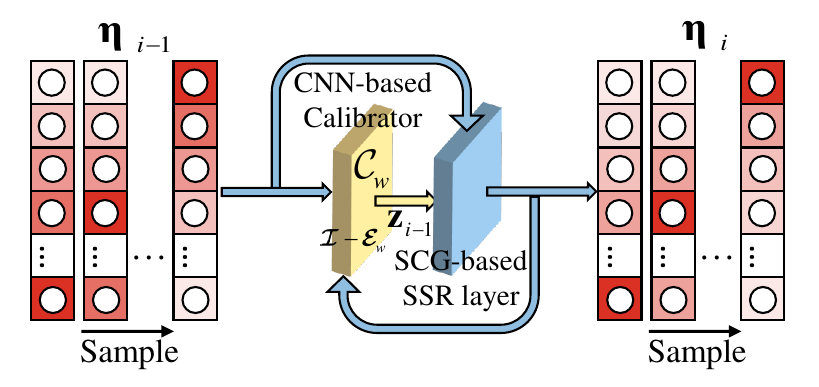}}
\vspace{-0.7em}
	\caption{Illustration of proposed iteration for MoD-DNN module. (a) Overall iteration in MoD-DNN module. (b) Alternative optimization between CNN-based calibrator and inverse problem-based SSR.}
\vspace{-0.8em}
	\label{iteration}
\end{figure*}
\vspace{-0.6em}

\subsection{Preprocessing}

\subsubsection{Coarray Spatial Spectrum}

Considering the discrete AoA set $\boldsymbol{\theta}$, the covariance matrix of $\mathbf{h}(k)$ in practical system is drived as
\begin{align}\label{covariance}
	\mathbf{R}={\rm{E}}\left[\mathbf{h}(k)\mathbf{h}^{\mathsf{H}}(k)\right] = \sum\nolimits_{l=1}^{L}\eta_l\tilde{\mathbf{a}}(\theta_l)\tilde{\mathbf{a}}^{\mathsf{H}}(\theta_l)+\sigma^2_{\rm{n}}\mathbf{I},
\end{align}
where $\eta_l$ represents the signal power on $\theta_l$, $\sigma_n^2$ is the power of the AWGN, and $\mathbf{I}$ denotes the identity matrix. Vectorizing the covariance matrix by stacking all columns of $\mathbf{R}$, we obtain the corresponding coarray signal  $\mathbf{y}=\tilde{\mathbf{A}}\boldsymbol{\eta}+\sigma^2_{\rm{n}}{\mathbf{e}}$, 
where $\tilde{\mathbf{A}} = [\tilde{\mathbf{A}}_1;\cdots;\tilde{\mathbf{A}}_M]$ is the equivalent coarray manifold and $\tilde{\mathbf{A}}_m=[\tilde{\mathbf{a}}(\theta_1)\tilde{\mathbf{a}}^{\mathsf{H}}(\theta_1)\mathbf{e}_m,\cdots,\tilde{\mathbf{a}}(\theta_L)\tilde{\mathbf{a}}^{\mathsf{H}}(\theta_L)\mathbf{e}_m]$. Notice that, $\tilde{\mathbf{A}}_m$ is not equal to the ideal case ${\mathbf{A}_m}=[{\mathbf{a}}(\theta_1){\mathbf{a}}^{\mathsf{H}}(\theta_1)\mathbf{e}_m,\cdots,{\mathbf{a}}(\theta_L){\mathbf{a}}^{\mathsf{H}}(\theta_L)\mathbf{e}_m]$, and $\boldsymbol{\eta}=[\eta_1,\cdots,\eta_L]$ can be considered as the sparse spatial spectrum to be estimated. ${\mathbf{e}}= \left[\mathbf{e}_1;\cdots;\mathbf{e}_M\right]$ is a transition vector, where $\mathbf{e}_m\in\mathbb{R}^{M}$ is a vector with the $m$-th element being 1 whereas the others are zero. Generally, the CSS can be estimated using digital beamforming as
\begin{align}\label{css}
	\hat{\boldsymbol{\eta}} = {\mathbf{A}}^{\mathsf{H}}\mathbf{y}\approx {\mathbf{A}}^{\mathsf{H}}\tilde{\mathbf{A}}\boldsymbol{\eta}=\tilde{\mathbf{P}}\boldsymbol{\eta}.
\end{align}
In this vain, the problem is equivalent to solve $\boldsymbol{\eta}$ from the observed spectrum $\hat{\boldsymbol{\eta}}$ with the projection matrix $\tilde{{\mathbf{P}}} = {\mathbf{A}}^{\mathsf{H}}{\tilde{\mathbf{A}}}$, and ${\mathbf{A}} = \left[{\mathbf{A}}_1;\cdots;{\mathbf{A}}_M\right]$.

\subsubsection{Inverse Problem}

At this moment, we have reformulated the AoA estimation problem to a inverse problem. However,  the projection matrix $\tilde{{\mathbf{P}}}$ usually differs from the ideal one ${{\mathbf{P}}}={\mathbf{A}}^{\mathsf{H}}{{\mathbf{A}}}$ and is unknown with existence of error matrix $\mathbf{\Gamma}$ as per (3). In fact, the $\gamma_{m}(\theta_l)$ is difficult to be estimated and compensated for by traditional methods. To estimate the AoA from the observed coarray signal in practical system, we propose a model-driven network with iterations between deep learning and model-driven signal processing methods.

\subsection{MoD-DNN Module}

\subsubsection{Iterative Optimization}

As shown in the overall iteration illustrated in Figure~\ref{iteration}(a), the AoA resolution of the CSS is relatively low, and the position of the peak is inaccurate. To enhance the AoA estimation performance, the reconstruction of the spatial spectrum is expressed as the following optimization problem
\begin{align}\label{optimization}
	\boldsymbol{\eta} = \mathop{\arg\min_{\mathbf{\boldsymbol{\eta}}}}\|\mathbf{P}\boldsymbol{\eta}-\hat{\boldsymbol{\eta}}\|_2^2+\lambda\|\mathcal{E}_\mathbf{w}(\boldsymbol{\eta})\|^2,
\end{align}
where $\mathcal{E}_{\mathbf{w}}$ is a regularization term denoting the combination of noise and hardware impairments, which can be learned by training a $\mathbf{w}$-weighted CNN. The regularization coefficient $\lambda$ is set as a trainable parameter as trade-off between calibration and reconstruction and can be fixed during the iterations.
As such, the regularization term  can be further expressed as $\mathcal{E}_{\mathbf{w}}(\boldsymbol{\eta}) = (\mathbf{I}-\mathcal{C}_{\mathbf{w}})(\boldsymbol{\eta}) = \boldsymbol{\eta} - \mathcal{C}_\mathbf{w}(\boldsymbol{\eta})$, where $\mathcal{C}_\mathbf{w}(\boldsymbol{\eta})$ is a calibrated version of $\boldsymbol{\eta}$ after the repairment of hardware impairments and noise. By utilizing the method of alternative optimization, the problem \eqref{optimization} is equivalent to the following iterations
\begin{subequations}\label{alternativeoptimization}
	\begin{equation}\label{iteration1}
		\mathbf{z}^i=\mathcal{C}_{\mathbf{w}}(\boldsymbol{\eta}^i),
	\end{equation}
	\begin{equation}\label{iteration2}
		\boldsymbol{\eta}^{i+1} = \mathop{\arg\min_{\mathbf{\boldsymbol{\eta}}}}\|\mathbf{P}\boldsymbol{\eta}-{\boldsymbol{\eta}}^{i}\|_2^2+\lambda\|\boldsymbol{\eta} - \mathbf{z}^i\|^2.
	\end{equation}
\end{subequations}
The illustration of the iteration is depicted in Figure~\ref{iteration}(b).

\subsubsection{SCG Algorithm}

Now we introduce the SCG algorithm used in the MoD-DNN module. We reconstruct the spatial spectrum $\boldsymbol{\eta}$ by solving the sub-problem \eqref{iteration2} using a forward model.  Conjugate gradient (CG) algorithm is usually utilized to solve such problem efficiently. All CG-steps have closed-form solutions which means no parameter needs extra training, and the gradients can be backpropagated from the CG sub-blocks in iterations. By integrating the CG algorithm, the MoD-DNN module has sub-blocks consisting of numerical optimization layer with low cost of training.

Although the CG algorithm can efficiently solve the problem, we recall that the original form of the inverse problem is $\mathbf{y} = \tilde{\mathbf{A}}\boldsymbol{\eta}$, where $\tilde{\mathbf{A}}$ is an over-complete coarray manifold corresponding to the discrete AoA set $\boldsymbol{\theta}$. This implies that the inverse problem is underdetermined which leads to wide beam-width for the solution using CG algorithm. To further enhance the reconstruction of CG method, we propose a sparsity-constrained CG algorithm for spectrum reconstruction. In order to fully utilize the sparsity of $\boldsymbol{{\eta}}$, we further regularized the optimization problem \eqref{iteration2} with a sparsity function $s(\boldsymbol{\eta})$ as
\begin{align}\label{optimization_scg}
	\boldsymbol{\eta}^{i+\!1}\!=\! \mathop{\arg\min_{\mathbf{\boldsymbol{\eta}}}}\|\mathbf{P}\boldsymbol{\eta}\!-\!{\boldsymbol{\eta}}^{i}\|_2^2+\lambda\|\boldsymbol{\eta} \!- \!\mathbf{z}^i\|^2+\mu\!\cdot\!s(\boldsymbol{\eta}),
\end{align}
where $\mu$ is the regularization coefficient with respect to sparsity constraint. Then, the minimization problem can be solved in an iterative method between CG solutions and sparsity modification. Without loss of generality, we can use reweighted zero attracting function as
\begin{align}
	s({\boldsymbol{\eta}}(n)) = {\rm{log}}\left(1+{\|\boldsymbol{\eta}(n)\|_1}/{\epsilon}\right),
\end{align}
where $\epsilon$ is an approximation parameter. The corresponding subgradient function in expressed as
\begin{align}
	\nabla^{s}s({\boldsymbol{\eta}}(n)) = \frac{{\rm{sgn}}(\boldsymbol{\eta}(n))}{1+\epsilon\|\boldsymbol{\eta}(n)\|_1},
\end{align}
where $\rm{sgn}(\cdot)$ is the signum function. We summarize the SCG-based spatial spectrum reconstruction (SSR) algorithm in  Algorithm \ref{alg_SCG}.

\begin{algorithm}[t]
	{\caption{SCG-based SSR algorithm.}
		\label{alg_SCG}
		\textbf{Input:} Coarray spectrum ${\boldsymbol{\eta}}^{i}$, calibrated spectrum $\mathbf{z}^{i}$, projection matrix $\mathbf{P}$, approximation parameter $\epsilon$.\\
		\textbf{Output:} Reconstructed spatial spectrum $\boldsymbol{\eta}^{i+1}$.\\
		\textbf{Initialize:} $\boldsymbol{\eta}^{i+1}(0)=\mathbf{0}$, $\mathbf{g}(0) = (\mathbf{P}+\lambda\mathbf{I})\boldsymbol{\eta}^{i+1}(0)-({\boldsymbol{\eta}}^{i}+\lambda\mathbf{z}^{i})$, $\mathbf{c}(0)=-\mathbf{g}(0)$, $\gamma_{\rm{CG}}$ and $N^{\rm{CGiter}}_{\rm{max}}$.
		\begin{algorithmic}[1]
			\For{$n=0$ to $N^{\text{iter}}_{\rm{max}}$}\\
			${\alpha}(n) = -\frac{\mathbf{g}^{\rm{T}}(n)\mathbf{c}(n)}{\mathbf{c}^{\rm{T}}(n)\mathbf{P}\mathbf{c}(n)}$\\
			$\boldsymbol{\eta}^{i+1}(n+1) = \boldsymbol{\eta}^{i+1}(n)+{\alpha}(n)\mathbf{c}(n)-\nabla^{s}s({\boldsymbol{\eta}^{i+1}}(n))$\\
			$\mathbf{g}(n+1) = (\mathbf{P}+\lambda\mathbf{I})\boldsymbol{\eta}^{i+1}(n)-({\boldsymbol{\eta}}^{i}+\lambda\mathbf{z}^{i})$\\
			${\beta}(n)=\frac{(\mathbf{g}(n+1)-\mathbf{g}(n))^{\rm{T}}\mathbf{g}(n+1)}{\mathbf{g^{\rm{T}}}(n)\mathbf{g}(n)}$\\
			$\mathbf{c}(n+1)=-\mathbf{g}(n+1)+{\beta}(n)\mathbf{c}(n)$
			\If{	
				$\boldsymbol{\eta}^{i+1}(n+1)-\boldsymbol{\eta}^{i+1}(n)<\gamma_{\rm{CG}}$}\\
			\qquad \quad Break;
			\EndIf
			\EndFor
	\end{algorithmic}}
\end{algorithm}

\subsubsection{Training of CNN-based Calibrator}

\begin{figure}[]
	\centering
	\includegraphics[width=1\linewidth]{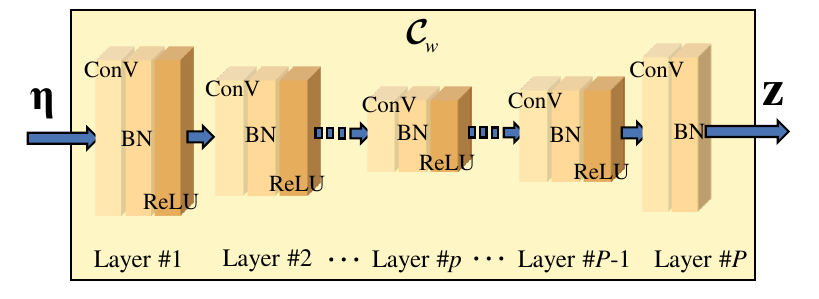}
	\vspace{-1.8em}
	\caption{Structure of CNN-based calibrator.}
	\label{calibrator}
	\vspace{-1em}
\end{figure}

In this subsection, we introduce the 1D-CNN structure which is used as the spectrum calibrator in \eqref{iteration1} to effectively calibrate non-linear hardware impairments widespread in real-world systems. Unlike the mapping relationships between array output and angle information in end-to-end learning method \cite{Yang2021Machine}, our approach transforms CSI (manifold format data) into image-format spatial spectrum. This conversion makes it well-suited for processing through CNNs. The structure of the 1D-CNN is given in Figure~\ref{calibrator}. Considering a training data set $\mathbb{D}=\left\{(\mathbf{y}(1),\boldsymbol{\eta}(1)),\cdots,(\mathbf{y}(D),\boldsymbol{\eta}(D)) \right\}$. The coarray signal $\mathbf{y}(d)$ can be obtained from the CSI-form receive signal using \eqref{covariance}, and the CSS $\boldsymbol{\hat{\eta}}(d)$ is generated using \eqref{css} and input to the CNN. Then, $P$ layers are used to calibrate the spatial spectrum. In this experiment, $4$ convolutional layers with convolutional kernel size of $32\times1$ are used to extract data features, and the number of kernels in each layer is $4$, $8$, $4$, and $1$, respectively. In addition, except for the activation function of the fourth layer, which uses Linear, the activation function of other layers uses ReLU. We select these hyper-parameters through ablation experiments using a controlled variable methodology. The input of CNN in the $i$-th iteration  is $\boldsymbol{\eta}^{i-1}$, and the calibrated spectrum is $\mathbf{z}^{i-1}$. Finally, $\mathbf{z}^{i-1}$ is input to the SCG algorithm to solve the sparse solution $\boldsymbol{\eta}^{i}$. When the iteration between the CNN and SCG modules is finished, the reconstructed spectrum $\boldsymbol{\eta}^{I}$ is the output of the framework. The goal of network training is minimizing the mean square error (MSE) loss function as
\begin{align}
	\mathcal{L} = \sum\nolimits_{d=1}^{D}\|\boldsymbol{\eta}^{I}(d)-\boldsymbol{\eta}(d)\|^2.
\end{align}
For the network training in PyTorch, the Adam optimizer and StepLR scheduler are utilized to optimize the MSE loss, which allows the computation of the gradients of the weights using backpropagation.

\section{Experimental Results}

In this section, we demonstrate the effectiveness of AoA estimation with existence of hardware impairments using the proposed MoD-DNN, especially in comparison with
multiple signal classification (MUSIC) algorithm, DeepMUSIC algorithm \cite{Elbir2020DeepMUSIC} and CNN \cite{Wang20212DCNN}. All novel datasets and codes introduced in this paper will be made publicly available upon publication of the paper with a license that allows free usage for research purposes.

\begin{figure*}[h]
	\vspace{-1em}
	\centering
	\subfloat[]{
		\includegraphics[width=0.3\linewidth]{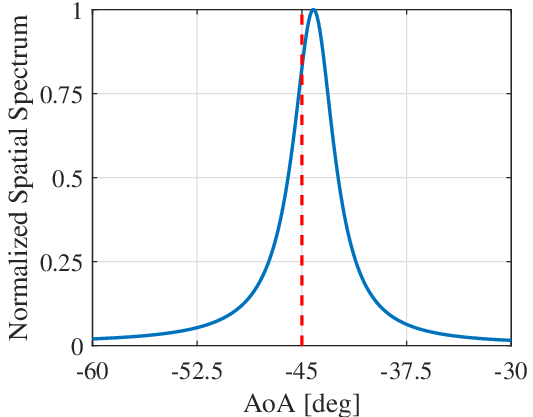}}\hfill
	\subfloat[]{
		\includegraphics[width=0.3\linewidth]{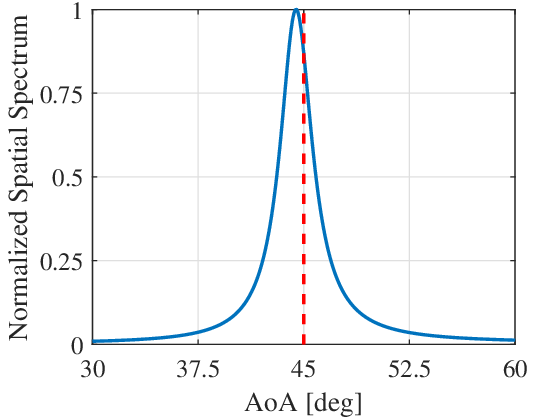}}\hfill
	\subfloat[]{
		\includegraphics[width=0.3\linewidth]{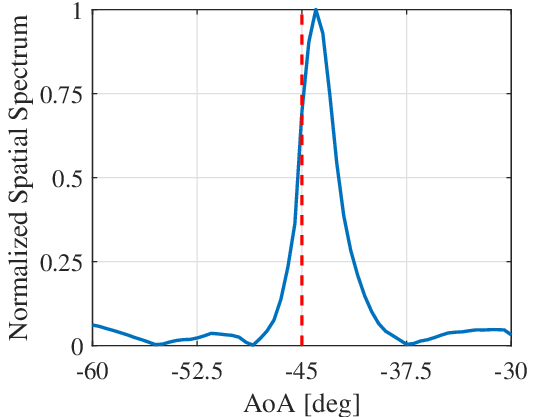}}
	\vspace{-1em}
	\hfill
	\subfloat[]{
		\includegraphics[width=0.3\linewidth]{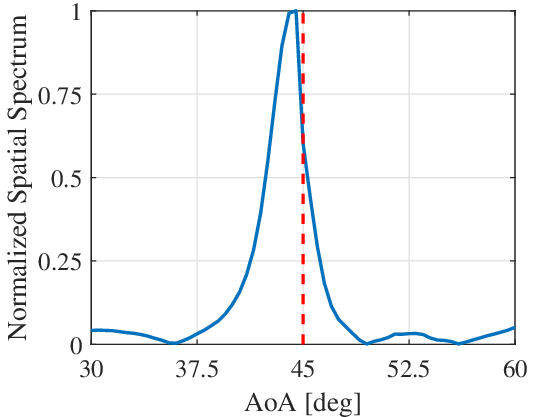}}\hfill
	\subfloat[]{
		\includegraphics[width=0.3\linewidth]{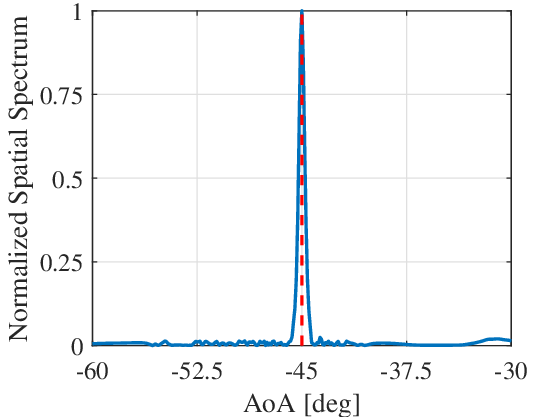}}\hfill
	\subfloat[]{
		\includegraphics[width=0.3\linewidth]{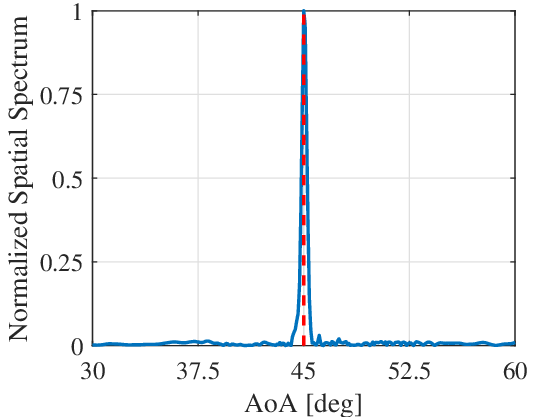}}
\vspace{-0.6em}
	\caption{Spatial spectrum results at different AoA. Red dashed lines denote the truth. (a) MUSIC, $-45^{\circ}$. (b) MUSIC, $45^{\circ}$. (c) DeepMUSIC, $-45^{\circ}$. (d) DeepMUSIC, $45^{\circ}$. (e) MoD-DNN, $-45^{\circ}$. (f) MoD-DNN, $45^{\circ}$.}
	\label{SS estimation}
	\graphicspath{{figure/}}
	\vspace{-1em}
\end{figure*}

\subsection{Numerical Simulations}

We consider a simulation scenario where a 5G gNB receives SRS signals transmitted by a user equipment (UE) in LOS condition. The CSI data is generated by the latest link-level 5G communications simulator \cite{Jia2023Link}. The simulator can incorporate customized hardware impairment functions and simulate channels in line with the 3GPP TR 38.901 standard. The indoor factory LoS channel at sub-6GHz working frequency (3GPP 38.901 InF LoS) is chosen throughout the simulations.

\subsubsection{Simulation Settings}

The central carrier frequency of the OFDM signal is $f_{\mathrm{c}}$, and uniform frequency increment is $\Delta f$. The distance of UE from 5G gNB is fixed at $R$. The AoA of UE varies in $\boldsymbol{\theta}$ with angular interval $\Delta \theta$. The parameter settings are listed in Table ~\ref{table:simulation}. Note that, $B$ and $P_{\text{UE}}$ respectively denote system bandwidth and power of UE. For each AoA, $45$ SRS symbols are generated with the existence of error matrix $\boldsymbol{\Gamma}$ by the simulator. As such, We extract $1201\times40 = 48040$ groups of data for network training and $1201\times5 = 6005$ for validation.

\begin{table}[b]
\vspace{-1em}
	\centering
	\begin{tabular}{c|c||c|c}
		\hline
		& & &\\[-8pt]
		Symbol&Value&Symbol&Value\\
		\hline
		& & &\\[-6pt]
		$f_{\text{c}}$&$4.8498\;\mathrm{GHz}$&$R$&$10\;\mathrm{m}$\\
		& & &\\[-6pt]
		$\Delta f$&$60\;\text{KHz}$&$P_{UE}$&$23\;\mathrm{dBm}$\\
		& & &\\[-6pt]
		$B$&$100\;\text{MHz}$&$\Delta \theta$&$0.1^{\circ}$\\
		& & &\\[-6pt]
		$M$&4&$\boldsymbol{\theta}$&$[-60^{\circ},60^{\circ}]$\\
		\hline
	\end{tabular}
	\caption{Simulation settings.}
	\label{table:simulation}
\end{table}

\subsubsection{Spatial Spectrum Results}

\begin{figure}[]
	\centering
	\subfloat[]{\includegraphics[width=0.96\linewidth,trim=12 0 32 16,clip]{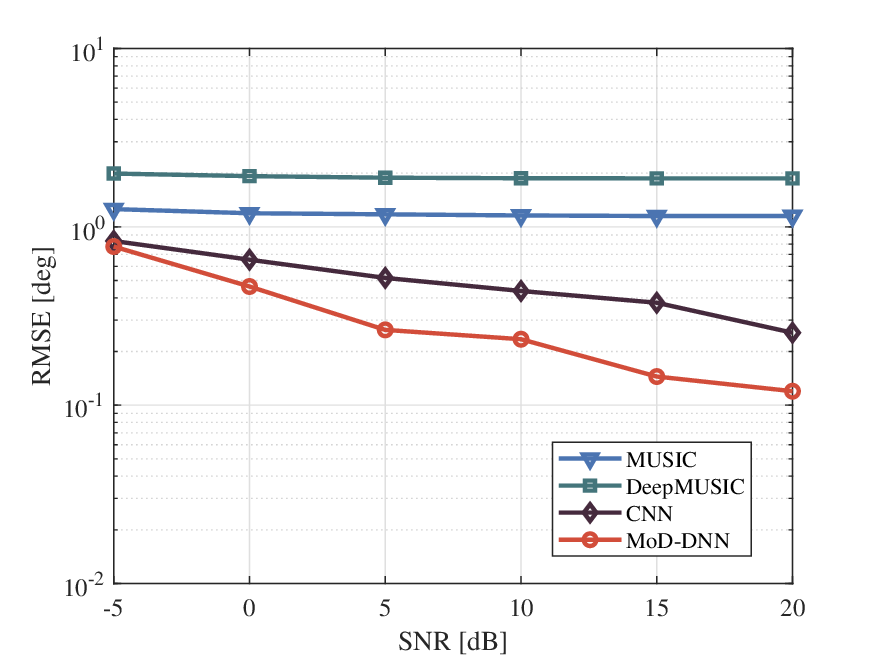}}\\
	\vspace{-0.6em}
	\subfloat[]{\includegraphics[width=0.96\linewidth,trim=12 0 32 16,clip]{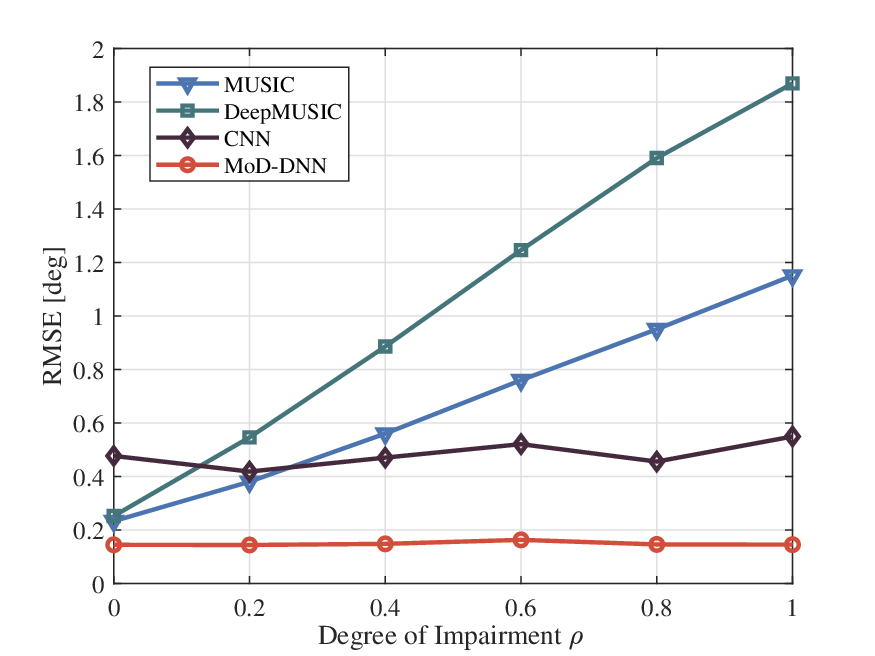}}\\
	\vspace{-0.6em}
	\subfloat[]{\includegraphics[width=0.96\linewidth,trim=12 0 32 16,clip]{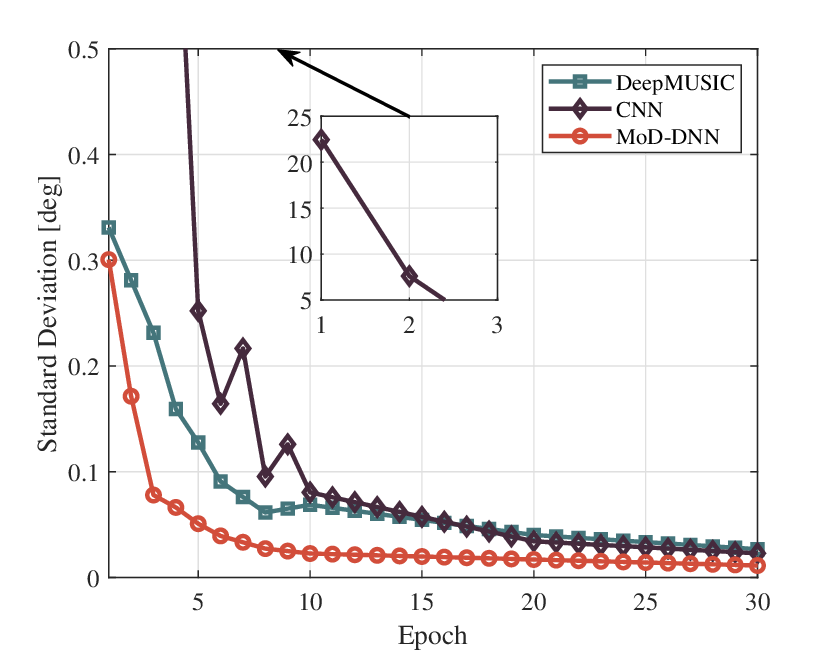}}
	\hfill
	\caption{Performance comparison of different methods. (a) RMSE versus SNR. (b) RMSE versus degree of impairment $\rho$. (c) Standard deviation versus epoch.}
	\label{RMSE_SD}
	\vspace{-1em}
\end{figure}

In the first set of simulations, the spatial spectrum results of MUSIC, DeepMUSIC and MoD-DNN at $-45^{\circ}$, $45^{\circ}$ are presented in Figure~\ref{SS estimation}, in which signal-to-noise ratio (SNR) is fixed to $10\;\mathrm{dB}$. It can be observed that MUSIC, DeepMUSIC and MoD-DNN can yield the spatial spectrum with a single peak which indicates the AoA of UE. However, due to the spectrum, the AoAs estimated by MUSIC and DeepMUSIC obviously deviate from the true value. As one of model-driven deep learning schemes, DeepMUSIC's performance is constrained without appropriate calibration for phase errors. Meanwhile, the spatial spectrum of the DeepMUSIC algorithm exhibits more fluctuations compared to the original MUSIC algorithm, which indicates a more severe impact of phase errors. As a comparison, the proposed MoD-DNN has the sharpest peak which accurately estimate the AoA.

\subsubsection{RMSE of AoA Estimation}

In order to further compare the estimation accuracy, we observe the RMSE of different algorithms versus SNR. As shown in Figure~\ref{RMSE_SD}(a), the performance of MUSIC and DeepMUSIC almost remain unchanged as SNR varies. This phenomenon indicates that these algorithm are statistically invalid  with the existence of hardware impairments. Different from that, the RMSEs of CNN and the proposed MoD-DNN both decrease while MoD-DNN consistently outperforms the CNN. Compared to the purely data-driven CNN, the proposed MoD-DNN can effectively utilize the knowledge of signal model. Accordingly, we observe that the advantage of MoD-DNN becomes more obvious in high SNR region. To validate the effectiveness for impairment calibration of the proposed method, we further compare the RMSEs for different cases of impairment. Specifically, we use a weight coefficient $\rho$ for error matrix $\boldsymbol{\Gamma}$ varies from $0$ to $1$ to simulate the degree of impairment. As illustrated in Figure~\ref{RMSE_SD}(b), DeepMUSIC and MUSIC perform well under low impairment condition. However, the performances become worse as $\rho$ increases. On the contrary, the performances of CNN and MoD-DNN only fluctuates slightly. Owing to the integration of NN and SCG, MoD-DNN successfully mitigates the influence of hardware impairments and outperforms the data-driven CNN.

\vspace{-0.7em}
\subsubsection{Convergence Speed of Algorithms}

In the final simulation,  we observe the standard deviation (SD) of loss versus epoch to evaluate the convergence speed of  different algorithms.  It is worth mentioning that we take the convergence value as the expectation  to calculate the standard deviation. As shown in Figure~\ref{RMSE_SD}(c), the SD of MoD-DNN converges rapidly compared to CNN and DeepMUSIC, which showcase the advantage of MoD-DNN with respect to training efficiency. So far, the effectiveness in terms of impairment calibration and estimation accuracy are validated by the numerical simulation.

\vspace{-0.7em}
\subsection{Experiment in an Anechoic Chamber}
\begin{figure}[h]
	\centering
	\includegraphics[width=0.9\linewidth]{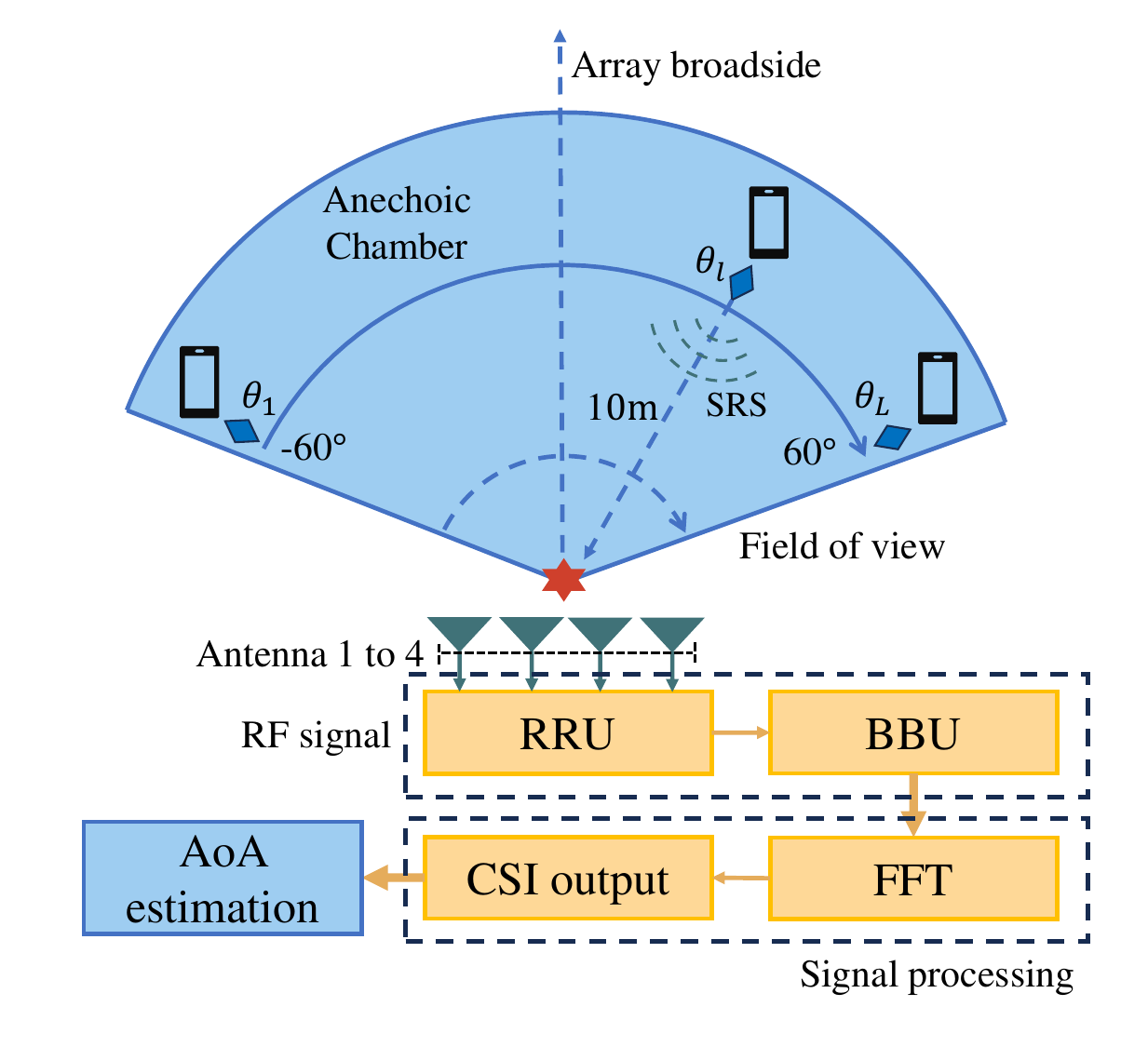}
\vspace{-1.2em}
	\caption{Setups for anechoic chamber experiments.}
	\label{fig:chamber}
\vspace{-1.3em}
\end{figure}

In this subsection, we compare the AoA estimation performance of different methods in an anechoic chamber. Experiments conducted in an anechoic chamber provide valuable insights as NLoS and multipath propagation are absent. In this controlled environment, hardware impairments are the primary challenges to the accuracy of AoA estimation. The experimental setups are illustrated in Figure~\ref{fig:chamber}.
\vspace{-1.1em}

\subsubsection{Experimental Settings}

We employs the UE and gNB for signal transmitting and receiving, respectively. The AoA of UE is rotated from $-60^{\circ}$ to  $60^{\circ}$ by a uniform angle interval of $1^{\circ}$ during the data acquisition process. For each AoA, $450$ SRS symbols are transmitted by the UE. Then, the sounded CSIs are collected by the 5G gNB, yielding $121\times450 = 54450$ groups of CSI data. We extract $121\times400 = 48400$ groups of data for network training and $121\times50 = 6050$ for validation.

\begin{figure}[h]
	\centering
	\subfloat[]{\includegraphics[width=0.8\linewidth,trim=10 0 20 10,clip]{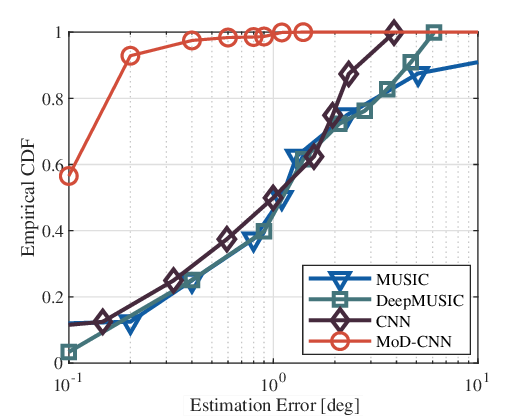}}\\\vspace{-0.7em}
	\subfloat[]{\includegraphics[width=0.95\linewidth,trim=25 0 45 45,clip]{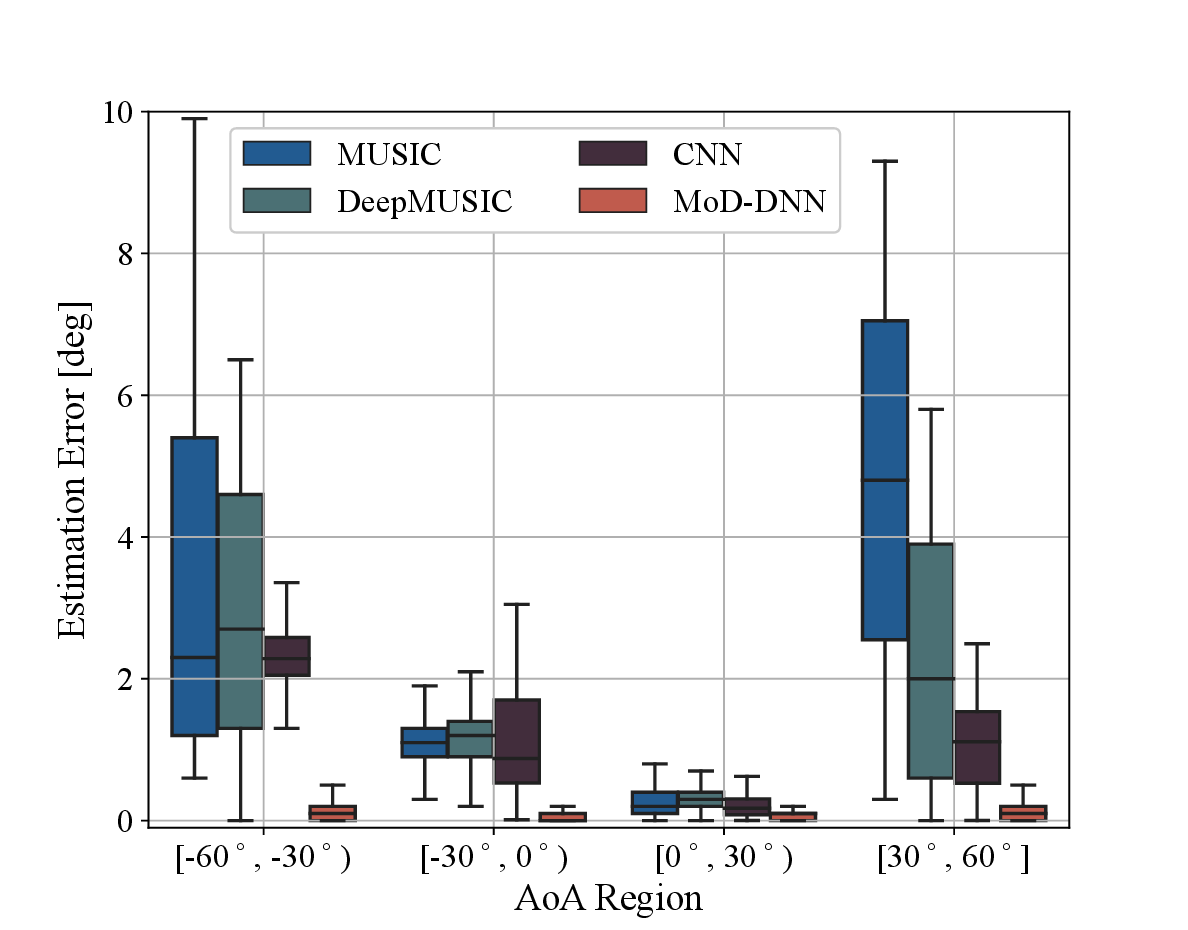}}\vspace{-0.7em}
	\caption{AoA estimation error performance of different methods. (a) CDF. (b)Boxplots for 4 subregions.}
	\label{fig:anechoic}\vspace{-1.5em}
\end{figure}

\subsubsection{CDF of AoA Estimation Error}

The cumulative distribution function (CDF) curves of AoA estimation error are provided in Figure~\ref{fig:anechoic}(a), which affirm the remarkable accuracy of the proposed MoD-DNN, outperforming all other methods. Particularly noteworthy is the MoD-DNN's capacity to achieve a minimum $95\%$ reduction in the 80th percentile of AoA estimation error , from around $3^{\circ}$ to $0.15^{\circ}$.

\subsubsection{Boxplots of AoA Estimation Error}

We proceed to evaluate the AoA estimation performance of various methods through boxplots portraying estimation errors across varying angular sectors. Specifically, the field of view is segmented into four subregions: $[-60^{\circ},-30^{\circ})$, $[-30^{\circ},-0^{\circ})$, $[0^{\circ},30^{\circ})$, and $[30^{\circ},60^{\circ}]$. The box's lower and upper boundaries correspond to the first quartile (Q1) and third quartile (Q3), respectively. The upper and lower whiskers respectively indicate the maximum and minimum estimation errors after removal of outliers (data with errors exceeding $1.5$ times Q3). The median is represented by a horizontal line positioned above the box.

As displayed in Fig,\;\ref{fig:anechoic}(b), within the angular regions of $[-30^\circ, -0^\circ)$ and $[0^\circ, 30^\circ)$, the estimation results of all algorithms exhibit relatively accurate performance, with median errors of approximately less than 1°. This outcome is attributed to the minimal impact of angular-dependent phase errors in these specific regions. As the AoA of the User Equipment (UE) extends to the angular regions of $[-60^\circ, -30^\circ)$ and $[30^\circ, 60^\circ]$, the severity of phase errors escalates. Consequently, the MUSIC and DeepMUSIC algorithms experience significant performance degradation. For these two algorithms, the maximum errors are approximately 10° and 6.5°, respectively, while the medians hover around 2°. The substantial fluctuation in phase errors also contributes to a high IQR for MUSIC and DeepMUSIC.In comparison, CNN and MoD-DNN are less affected by these conditions. Remarkably, the proposed MoD-DNN algorithm demonstrates superior performance across all assessed indicators represented by the box plots, encompassing maximum error, median error, and IQR in all angular regions. This phenomenon indicates a robust estimation performance of MoD-DNN and validates the effectiveness of error calibration for MoD-DNN.
\subsubsection{Computational Complexity}
Lastly, the computational efficiency of MUSIC, DeepMUSIC, CNN, and MoD-DNN are compared. The running times are all recorded by the timer of PyTorch on a PC with Intel(R) Core(TM) i7-1065G7 CPU and a 16-GB RAM. The result is shown in TABLE.\;\ref{table:Complexity} while the training time is mentioned in hours, whereas the testing time is mentioned in milliseconds.

In comparison to algorithms generating spatial spectra as output, the CNN algorithm boasts a notably shorter training duration due to its direct output of low-dimensional AoA values. In the case of the MoD-DNN algorithm and the DeepMUSIC algorithm, the former capitalizes on the weight-sharing strategy within the CNN module and the closed-form solutions of the SCG algorithm. Consequently, the training parameters number of MoD-DNN is fewer than that of DeepMUSIC, resulting in a clear advantage in training time. In terms of testing time, it's clear that the computational time for all algorithms stays within the millisecond range, thus ensuring efficient real-time implementation.
\begin{table}[htbp!]
	\centering
	\begin{tabular}{c c c c c}
		\hline
		& & & & \\[-6pt]
		&MUSIC&DeepMUSIC&CNN&\textbf{Ours}\\
		\hline
		& & & & \\[-6pt]
		training [h]&-& 30.8 & 1.6 & \textbf{15.6} \\
		
		& & & & \\[-6pt]
		testing [ms]&9.4 &31.6 &1.8 &\textbf{14.9}\\
		\hline
	\end{tabular}
	\caption{Training and testing times.}
	\label{table:Complexity}
\end{table}

\vspace{-1em}
\section{Conclusion}

This study addresses AoA estimation in the presence of hardware impairments. We reformulate direction finding as spatial spectrum reconstruction through an inverse problem formulation. The iterative optimization between CNN and SCG enhances spatial spectrum calibration, leading to improved estimation precision. Simulation and experimental findings confirm the superiority of our proposed framework over both pure model-driven and data-driven approaches. Our insights into hybrid AI-and-model-driven methodologies hold promise for mitigating radio-frequency hardware-related challenges and enhancing the overall system reliability and efficiency for the broader spectrum of wireless tasks.

\section*{Acknowledgements}
This work was supported in part by the National Natural Science Foundation of China under Grant No. 62225107, the Natural Science Foundation on Frontier Leading Technology Basic Research Project of Jiangsu under Grant No. BK20222001, and the Fundamental Research Funds for the Central Universities under Grant No. 2242022k60002.

\newpage
\bibliography{aaai24}

\end{document}